\documentclass[envcountsame,envcountsect,undated,nolinenumbers]{lnthi}

\usepackage{epsf}

\title{An Optimal Choice Dictionary}

\author{Torben Hagerup}
\institute{Institut f\"ur Informatik, Universit\"at Augsburg, 86135
Augsburg, Germany
\email{hagerup@informatik.uni-augsburg.de}}

\setpagewiselinenumbers

\def\Tvn#1{\hbox{\textit{#1\/}}}
\def\Tfloor#1{\lfloor #1\rfloor}
\def\Tceil#1{\lceil #1\rceil}
\def\TbbbN{\mathbb{N}}
\def\wik{\widetilde{k}}

\begin{document}
\overfullrule=5pt
\maketitle{}

\begin{abstract}
A choice dictionary is a
data structure that can be initialized
with a parameter $n\in\TbbbN=\{1,2,\ldots\}$
and subsequently maintains an initially empty subset $S$
of $\{1,\ldots,n\}$ under insertion, deletion, membership queries
and an operation \Tvn{choice} that returns
an arbitrary element of $S$.
The choice dictionary is fundamental
in space-efficient computing and has numerous applications.
The best previous choice dictionary can be initialized
with $n$ and a second parameter $t\in\TbbbN$ in constant time
and subsequently executes all
operations in $O(t)$ time and occupies
$n+O(n({t/w})^t+\log n)$ bits on a word
RAM with a word length of $w=\Omega(\log n)$ bits.
We describe a new choice dictionary that,
following a constant-time initialization,
executes all operations in constant time
and, in addition to the space
needed to store~$n$, occupies only $n+1$ bits,
which is shown to be optimal if $w=o(n)$.
Allowed $\Tceil{\log_2(n+1)}$ bits of additional space,
the new data structure also supports iteration over
the set $S$ in constant time per element.

\end{abstract}

\pagestyle{plain}
\thispagestyle{plain}

\section{Introduction}
\label{sec:intro}
Following similar earlier
definitions~\cite{BriT93,ElmHK15} and concurrently
with that of \cite{BanCR16},
the choice-dictionary data type was introduced
by Hagerup and Kammer~\cite{HagK16} as a basic tool in
space-efficient computing and
is known to have
numerous applications
\cite{BanCR16,ElmHK15,HagK16,HagKL17,KamKL16}.
Its precise characterization is as follows:

\begin{definition}
\label{def:choice}
A \emph{choice dictionary} is a data type that
can be initialized with an arbitrary integer
$n\in\TbbbN=\{1,2,\ldots\}$,
subsequently maintains
an initially empty subset $S$ of $U=\{1,\ldots,n\}$
and supports the following operations,
whose preconditions are stated in
parentheses:

\begin{tabbing}
\quad\=\hskip 2.4cm\=\hskip 2cm\=\kill
\>$\Tvn{insert}(\ell)$\>($\ell\in U$):\>
Replaces $S$ by $S\cup\{\ell\}$.\\
\>$\Tvn{delete}(\ell)$\>($\ell\in U$):\>
Replaces $S$ by $S\setminus\{\ell\}$.\\
\>$\Tvn{contains}(\ell)$\>($\ell\in U$):\>
Returns 1 if $\ell\in S$, 0 otherwise.\\
\>$\Tvn{choice}$:\>
\>Returns an (arbitrary) element of $S$
if $S\not=\emptyset$, 0 otherwise.
\end{tabbing}
\end{definition}

As is common and convenient, we use the term
``choice dictionary'' also to denote data structures
that implement the choice-dictionary data type.
Following the initialization of a choice dictionary
$D$ with an integer~$n$,
we call (the constant) $n$ the \emph{universe size}
of $D$ and (the variable) $S$ its
\emph{client set}.
If a choice dictionary $D$ can operate only if
given access to $n$ (stored outside of~$D$),
we say that $D$ is \emph{externally sized}.
Otherwise, for emphasis, we may call $D$
\emph{self-contained}.

Our model of computation is a word RAM~\cite{AngV79,Hag98}
with a word length of $w\in\TbbbN$ bits, where we assume that $w$ is
large enough to allow all memory words in use to be addressed.
As part of ensuring this,
in the discussion of a choice dictionary
with universe size~$n$ we always assume that $w>\log_2\! n$.
The word RAM has constant-time operations
for addition, subtraction and multiplication
modulo $2^w$, division with truncation
($(x,y)\mapsto\Tfloor{{x/y}}$ for $y>0$),
left shift modulo $2^w$
($(x,y)\mapsto (x\ll y)\bmod 2^w$,
where $x\ll y=x\cdot 2^y$),
right shift
($(x,y)\mapsto x\gg y=\Tfloor{{x/{2^y}}}$),
and bitwise Boolean operations
($\textsc{and}$, $\textsc{or}$ and $\textsc{xor}$
(exclusive or)).
We also assume a constant-time operation to
load an integer that deviates from $\sqrt{w}$
by at most a constant factor---this enables the
proof of Lemma~\ref{lem:log}.

The best previous choice dictionary~\cite[Theorem~7.6]{HagK16}
can be initialized
with a universe size $n$ and a second parameter $t\in\TbbbN$
in constant time
and subsequently executes all
operations in $O(t)$ time and occupies
$n+O(n({t/w})^t+\log n)$ bits.
Let us call a choice dictionary \emph{atomic}
if it executes all operations including the
initialization in constant time.
Then, for every constant $t\in\TbbbN$,
the result just cited implies the existence of
an atomic choice dictionary that occupies
$n+O(n/{w^t}+\log n)$ bits when
initialized for universe size~$n$.
Here we describe an externally sized atomic
choice dictionary that needs just $n+1$ bits,
which is optimal if $w=o(n)$.
The optimality of the bound of $n+1$ bits
follows from a
simple argument of~\cite{HagK17,KatG17}:
Because the client set $S$ of a choice dictionary
with universe size $n$ can be in $2^n$ different
states, any two of which can be distinguished
via calls of $\Tvn{contains}$, if the choice dictionary
uses only $n$ bits it must represent each
possible state of $S$ through a unique bit pattern.
Since $S$ is in one particular state following
the initialization, the latter must force each
of $n$ bits to a specific value, which needs
$\Omega({n/w})$ time.

In addition to being more space-efficient than
all earlier choice dictionaries, the new data structure
is also significantly simpler than its best
predecessors, and in an actual
implementation its operations are likely to be
faster by a constant factor.
The new choice dictionary owes much to a recent
data structure of Katoh and Goto~\cite{KatG17}
that implements so-called \emph{initializable arrays}.
A connection between choice dictionaries and
initializable arrays was first noted by
Hagerup and Kammer~\cite{HagK17}, who observed
that the \emph{light-path technique}, invented in~\cite{HagK16}
in the context of choice dictionaries, also yields
initializable arrays better than those
known at the time.
Here we show that an ingenious and elegant data
representation devised by Katoh and Goto, slightly
modified and used together with additional operations,
yields a choice dictionary that leaves little
to be desired.
Our main result, formulated in Theorem~\ref{thm:main}
below, can be expressed informally as follows:
At the price of one additional bit, a bit vector can be
augmented with the operations
``clear all'' and ``locate a \texttt{1}''.

\begin{theorem}
\label{thm:main}
There is an externally sized atomic
choice dictionary that,
when initialized for universe size~$n$,
occupies $n+1$ bits
in each
quiescent state (i.e., between operations) and
needs $O(w)$ additional bits of transient space
during the execution of each operation.
\end{theorem}

\section{A Simple Reduction}
\label{sec:reduction}

For $n\in\TbbbN$, the \emph{bit-vector representation}
over $U=\{1,\ldots,n\}$ of a subset $S$ of $U$
is the sequence $(d_1,\ldots,d_n)$ of $n$
bits with $d_\ell=\texttt{1}\Leftrightarrow\ell\in S$,
for $\ell=1,\ldots,n$, or its obvious
layout in $n$ consecutive bits in memory.
If we represent the client set of a choice
dictionary with universe size $n$ via its
bit-vector representation~$B$, the choice-dictionary
operations translate into the reading and writing
of individual bits in $B$ and the operation
\Tvn{choice}, which now returns the position of
a nonzero bit in~$B$ (0 if all bits in $B$
are~\texttt{0}).
It is trivial to carry out all operations other
than initialization and \Tvn{choice}
in constant time.
In the special case $n=O(w)$, the latter
operations can also be supported in constant time.
This is a consequence of the following lemma.

\begin{lemma}[\cite{FreW93,HagK16}]
\label{lem:log}
Given a nonzero integer $\sum_{j=0}^{w-1}2^j d_j$,
where $d_j\in\{0,1\}$ for $j=0,\ldots,w-1$,
constant time suffices to compute $\max J$
and $\min J$, where $J=\{j\in\{0,\ldots,w-1\}\mid d_j=1\}$.
\end{lemma}

We use the
externally sized atomic choice dictionary for universe
sizes of $O(w)$ implied by these considerations
to handle the few bits left over when we divide
a bit-vector representation of $n$ bits into
pieces of a fixed size.
The details are as follows:

Let $b$ be a positive integer that can be computed
from $w$ and $n$ in constant time using $O(w)$ bits
(and therefore need not be stored) and that
satisfies $b\ge\log_2\! n$, but $b=O(w)$.
In order to realize an
externally sized choice dictionary $D$ with
universe size~$n$ and client set~$S$, partition the
bit-vector representation $B$ of $S$ into
$N=\Tfloor{n/{(2 b)}}$
segments $B_1,\ldots,B_N$
of exactly $2 b$ bits each, with
$n'=n\bmod(2 b)$ bits left over.
If $n'\not=0$, maintain (the set
corresponding to) the last $n'$
bits of $B$ in an externally sized
atomic choice dictionary $D_2$
realized as discussed above.
Assume without loss of generality that $N\ge 1$.
The following lemma is proved in the
remainder of the paper:

\begin{lemma}
\label{lem:main}
There is a data structure that,
if given access to $b$ and $N$,
can be initialized
in constant time and subsequently occupies $2 b N+1$ bits and
maintains a sequence
$(a_1,\ldots,a_N)\in\{0,\ldots,2^{2 b}-1\}^N$,
initially $(0,\ldots,0)$, under
the following operations, all of which
execute in constant time:
$\Tvn{read}(i)$ ($i\in\{1,\ldots,N\}$), which
returns $a_i$;
$\Tvn{write}(i,x)$ ($i\in\{1,\ldots,N\}$ and
$x\in\{0,\ldots,2^{2 b}-1\}$), which 
sets $a_i$ to $x$;
and \Tvn{nonzero},
which returns an $i\in\{1,\ldots,N\}$
with $a_i\not=0$
if there is such an $i$, and 0 otherwise.
\end{lemma}

For $i=1,\ldots,N$, view $B_i$ as the binary
representation of an integer and maintain
that integer as~$a_i$ in an instance of
the data structure of Lemma~\ref{lem:main}.
This yields an externally sized
atomic choice dictionary
$D_1$ for the first $2 b N$ bits of $B$:
To carry out \Tvn{insert}, \Tvn{delete}
or \Tvn{contains}, update or inspect the
relevant bit in one of $a_1,\ldots,a_N$,
and to execute \Tvn{choice}, call \Tvn{nonzero}
and, if the return value $i$ is positive,
apply an algorithm of Lemma~\ref{lem:log} to~$a_i$.
It is obvious how to realize the full
choice dictionary $D$ through a combination
of $D_1$ and $D_2$.
The only nontrivial case is that of
the operation \Tvn{choice}.
To execute \Tvn{choice} in $D$, first call
\Tvn{choice} in $D_1$ (say).
If the return value is positive, it is a suitable
return value for the parent call.
Otherwise call \Tvn{choice} in $D_2$,
increase the return value by $2 b N$ if it is positive,
and return the resulting number.
$D$ is atomic because $D_1$ and $D_2$ are, and
the total number of bits used by $D$
is $2 b N+1+n'=n+1$.
Theorem~\ref{thm:main} follows.

\section{The Main Construction}
\label{sec:main}

In this section we prove Lemma~\ref{lem:main},
except that we relax the space bound by
allowing $2 b N+w$ bits instead of $2 b N+1$ bits.

\subsection{The Storage Scheme}

This subsection describes how the sequence
$(a_1,\ldots,a_N)$ is represented in memory in
$2 b N+w$ bits.
Most of the available memory stores an array $A$
of $N$ cells $A[1],\ldots,A[N]$ of $2 b$ bits each.
In addition, a $w$-bit word is used to hold
an integer $k\in\{0,\ldots,N\}$ best thought of as a
``barrier'' that divides $V=\{1,\ldots,N\}$
into a part to the left of the barrier, $\{1,\ldots,k\}$,
and a part to its right, $\{k+1,\ldots,N\}$.
We often consider a $(2 b)$-bit quantity
$x$ to consist of a \emph{lower half}, denoted by
$\underline{x}$ and composed of the $b$
least significant bits of $x$
(i.e., $\underline{x}=x\,\mathbin{\textsc{and}}\,(2^b-1)$)
and an \emph{upper half},
$\overline{x}=x\gg b$, and we may
write $x=(\underline{x},\overline{x})$.
A central idea is that the upper halves of
$A[1],\ldots,A[N]$ are used to implement a
matching on $V$
according to the following convention:
Elements $i$ and $j$ of $V$ are matched exactly if
$\overline{A[i]}=j$, $\overline{A[j]}=i$,
and precisely one of $i$ and $j$ lies to the
left of the barrier, i.e., $i\le k<j$
or $j\le k<i$.
In this case we call $j$ the \emph{mate} of
$i$ and vice versa.
The assumption $b\ge\log_2\! n$ ensures that
the upper half of each cell in~$A$ can hold an arbitrary
element of~$V$.
A function that inputs an element $i$ of $V$
and returns the mate of $i$ if $i$ is matched
and $i$ itself if not is easily coded as follows:

\begin{tabbing}
\quad\=\quad\=\quad\=\quad\=\kill
\Tvn{mate}$(i)$:\\
\>$i':=\overline{A[i]}$;\\
\>\textbf{if} $(1\le i\le k<i'\le N$ or
$1\le i'\le k<i\le N)$ and $\overline{A[i']}=i$
\textbf{then return} $i'$;\\
\>\textbf{return} $i$;
\end{tabbing}

\noindent
For all $i\in V$, call $i$ \emph{strong}
if $i$ is matched and $i\le k$ or
$i$ is unmatched and $i>k$,
and call $i$ \emph{weak} if it is not strong.
The integers $A[1],\ldots,A[N]$ and $k$
represent the sequence $(a_1,\ldots,a_N)$
according to the following storage invariant:
For all $i\in V$,

\begin{itemize}
\item
$a_i=0$ exactly if $i$ is weak;
\item
if $i$ is strong and $i>k$, then $a_i=A[i]$;
\item
if $i$ is strong and $i\le k$, then
$a_i=(\underline{A[i]},\underline{A[\Tvn{mate}(i)]})$.
\end{itemize}

\begin{figure}
\begin{center}
\epsffile{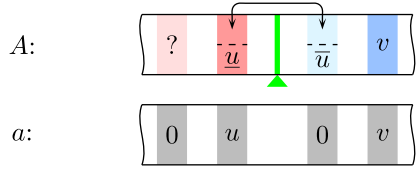}
\end{center}
\caption{The storage scheme.
Above: The array $A$.
Below: The sequence $a$ represented by $A$.}
\label{fig:scheme}
\end{figure}

The storage invariant is illustrated in Fig.~\ref{fig:scheme}.
The following drawing conventions are used here
and in subsequent figures:
The barrier is shown as a thick vertical line segment
with a triangular base.
Each pair of mates is connected with a double arrow,
and a cell $A[i]$ of $A$ is shown in a
darker hue if $i$ is strong.
A question mark indicates an entry that can be
completely arbitrary, except that it may not give
rise to a matching edge, and the upper and lower halves
of some cells of $A$ are shown separated by
a dashed line segment.

\subsection{The Easy Operations}

The data structure is initialized by setting $k=N$,
i.e., by placing the barrier at the right end.
Then the matching is empty, and all elements of $V$
are to the left of the barrier and weak.
Thus the initial value of $(a_1,\ldots,a_N)$
is $(0,\ldots,0)$, as required.
The implementation of $\Tvn{read}$
closely reflects
the storage invariant:

\begin{tabbing}
\quad\=\quad\=\quad\=\quad\=\kill
\Tvn{read}$(i)$:\\
\>\textbf{if} $\Tvn{mate}(i)\le k$ \textbf{then return} 0;
 $(*$ $i$ is weak exactly if $\Tvn{mate}(i)\le k$ $*)$\\
\>\textbf{if} $i>k$ \textbf{then return}
$A[i]$; \textbf{else return}
$(\underline{A[i]},\underline{A[\Tvn{mate}(i)]})$;
\end{tabbing}

\noindent
The code for \Tvn{nonzero} is short but a
little tricky:

\begin{tabbing}
\quad\=\quad\=\quad\=\quad\=\kill
\Tvn{nonzero}:\\
\>\textbf{if} $k=N$ \textbf{then return} 0;\\
\>\textbf{return} $\Tvn{mate}(N)$;
\end{tabbing}

The implementation of $\Tvn{write}(i,x)$
is easy if $i$ is weak and
$x=0$ (then nothing needs to be done) or
$i$ is strong and $x\not=0$.
In the latter case the procedure \Tvn{simple\_write}
shown below can be used.
The only point worth noting is that writing
to $A[i]$ when $i$ is strong and $i>k$
may create a spurious matching edge that
must be eliminated.

\begin{tabbing}
\quad\=\quad\=\quad\=\quad\=\kill
\Tvn{simple\_write}$(i,x)$:\\
\>\textbf{if} $i\le k$ \textbf{then}
$(\underline{A[i]},
\underline{A[\Tvn{mate}(i)]}):=
 (\underline{x},\overline{x})$;\\
\>\textbf{else}\\
\>\>$A[i]:=x$;\\
\>\>$i':=\Tvn{mate}(i)$;\\
\>\>\textbf{if} $i'\not=i$ \textbf{then} $\overline{A[i']}:=i'$;
 $(*$ eliminate a spurious matching edge $*)$
\end{tabbing}

\subsection{Insertion and Deletion}

The remaining, more complicated, operations of
the form $\Tvn{write}(i,x)$ are those in which
$a_i$ is changed from zero to
a nonzero value---call such an operation an
\emph{insertion}---or vice versa---a \emph{deletion}.
When the data structure under development is used
to realize a choice dictionary, insertions and
deletions
are triggered by (certain)
insertions and
deletions, respectively, executed on the choice dictionary.
Insertions and deletions
are the operations that change the
barrier and usually the matching.
In fact, $k$ is decreased by~1 in every
insertion and increased by~1 in every deletion,
so $k$ is always the number of
$i\in V$ with $a_i=0$.

\begin{figure}
\begin{center}
\epsffile{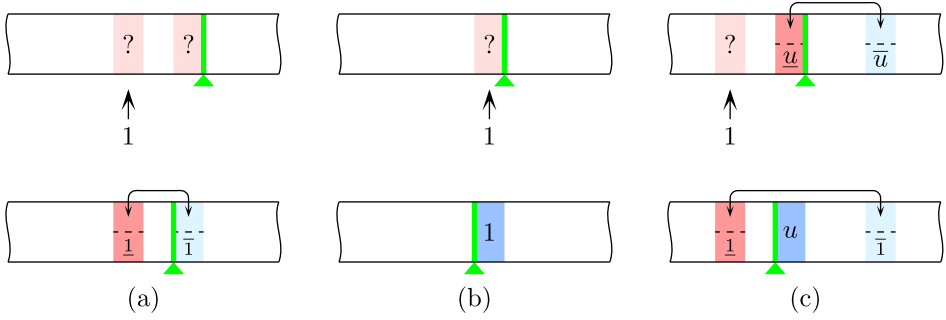}
\caption{Insertion to the left of the barrier.}
\label{fig:2}
\end{center}
\end{figure}

\begin{figure}
\begin{center}
\epsffile{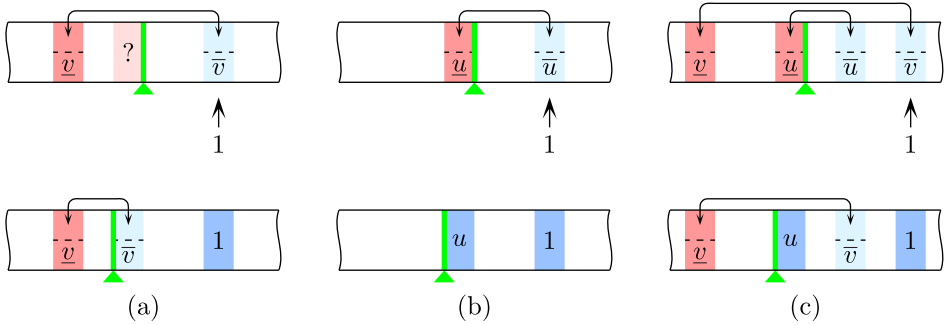}
\caption{Insertion to the right of the barrier.}
\label{fig:3}
\end{center}
\end{figure}

\begin{figure}
\begin{center}
\epsffile{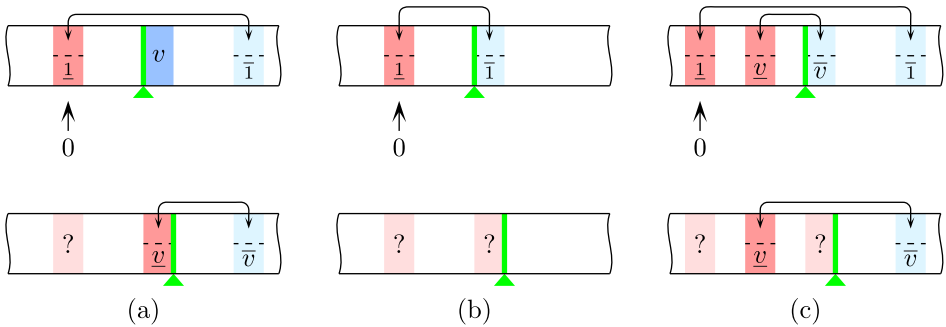}
\caption{Deletion to the left of the barrier.}
\label{fig:4}
\end{center}
\end{figure}

\begin{figure}
\begin{center}
\epsffile{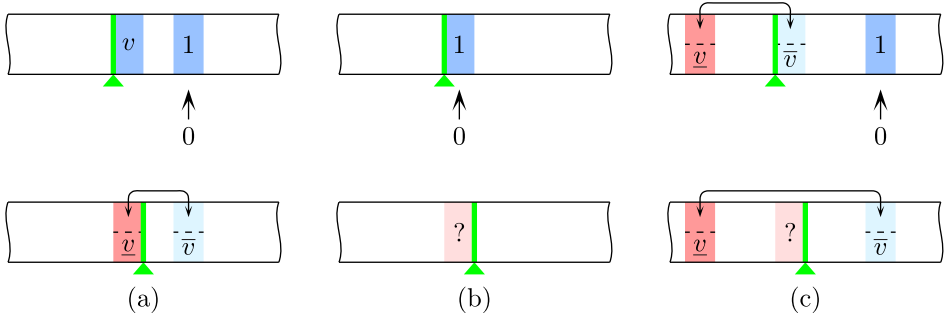}
\caption{Deletion to the right of the barrier.}
\label{fig:5}
\end{center}
\end{figure}

The various different forms that an insertion
may take are illustrated in Figs.\
\ref{fig:2} and~\ref{fig:3}.
The situation before the insertion is always
shown above the situation after the insertion.
A ``1'' outside of the ``stripes'' indicates
the position of an insertion and symbolizes
the nonzero value to be written, while a ``1'' inside
the stripes symbolizes that value after
it has been written.
The various forms of a deletion are illustrated
in Figs.\ \ref{fig:4} and~\ref{fig:5}.
Here a ``0'' indicates the position of a deletion,
while a ``1'' symbolizes the nonzero value that
is to be replaced by zero.

There are many somewhat different
cases, but for each it is easy to
see that the storage invariant
is preserved and that
the sequence $(a_1,\ldots,a_N)$
changes as required.
It is also easy to turn the figures into a
\Tvn{write} procedure that branches
into as many cases.
Here we propose the
following realization of \Tvn{write}
that is terser, but needs a more
careful justification.

\begin{tabbing}
\quad\=\quad\=\quad\=\quad\=\kill
\Tvn{write}$(i,x)$:\\
\>$x_0:=\Tvn{read}(i)$; $(*$ the value to be replaced by $x$ $*)$\\
\>$i':=\Tvn{mate}(i)$;\\
\>\texttt{if} $x\not=0$ \texttt{then}\\
\>\>\texttt{if} $x_0=0$ \texttt{then} $(*$ an insertion $*)$\\
\>\>\>$k':=\Tvn{mate}(k)$; $(*$ $\wik=k$ will cross the barrier $*)$\\
\>\>\>$u:=\Tvn{read}(k)$; $(*$ save $a_{\wik}$ $*)$ \\
\>\>\>$k:=k-1$; $(*$ move the barrier left $*)$\\
\>\>\>\Tvn{simple\_write}$(k+1,u)$;
 $(*$ reestablish the value of $a_{\wik}$ $*)$\\
\>\>\>\textbf{if} $i\not=k'$ \texttt{then} $\{$
$\overline{A[i']}:=k'$; $\overline{A[k']}:=i'$;
 $\underline{A[k']}:=\underline{A[i]}$; $\}$
 $(*$ match $i'$ and $k'$ $*)$\\
\>\>\Tvn{simple\_write}$(i,x)$; $(*$ $i$ was or has been made strong $*)$\\
\>\textbf{else} $(*$ $x=0$ $*)$\\
\>\>\texttt{if} $x_0\not=0$ \texttt{then} $(*$ a deletion $*)$\\
\>\>\>$k':=\Tvn{mate}(k+1)$; $(*$ $\wik=k+1$ will cross the barrier $*)$\\
\>\>\>$v:=\Tvn{read}(k')$; $(*$ save $a_{k'}$ $*)$ \\
\>\>\>$k:=k+1$; $(*$ move the barrier right $*)$\\
\>\>\>$\overline{A[i']}:=k'$; $\overline{A[k']}:=i'$;
 $(*$ match $i'$ and $k'$ $*)$\\
\>\>\>\texttt{if} $k'\not=i$ \texttt{then}
 $\Tvn{simple\_write}(k',v)$; $(*$ reestablish the
 value of $a_{k'}$ $*)$
\end{tabbing}

To see the correctness of the procedure
\Tvn{write} shown above,
consider a call $\Tvn{write}(i,x)$ and assume that
it gives rise to an insertion or a deletion,
since in the remaining cases the procedure is
easily seen to perform correctly.
Let $k_0$ be the value of $k$ (immediately)
before the call.
Because the call changes the value of $k$ by~1,
a single element $\wik$ of $V$
\emph{crosses} the barrier, i.e., is to the left
of the barrier before or after the call, but not both.
In the case of an insertion, $\wik=k_0$;
in that of a deletion, $\wik=k_0+1$.

Assume that $i$ does not cross the barrier,
i.e., that $i\not=\wik$.
Because the call changes $a_i$ from zero to a
nonzero value or vice versa, $i$ must change its
matching status, i.e., be matched before or after
the call, but not both.
In detail, if $i$ is matched before the call,
its mate at that time, if different from $\wik$,
must find a new mate, which automatically
leaves $i$ unmatched.
If $i$ is unmatched before the call, $i$ itself
must find a mate.
We can unify the two cases by saying that if
$i'=\Tvn{mate}(i)$ (evaluated before the call
under consideration has changed $k$ and~$A$)
is not $\wik$, then $i'$ must find a (new) mate.
If $i'\not=\wik$, moreover, $i'$ is to the
left of the barrier in the case of an insertion
and to the right of it in the case of a deletion.

Assume now that the call does not change
$a_{\wik}$, i.e., that $\wik\not=i$.
Then, because $\wik$ crosses the barrier,
it must also change its matching status:
If $\wik$ is matched before the call,
its mate at that time, if different from $i$,
must find a new mate, and otherwise $\wik$
itself must find a mate.
As above, this can be expressed by saying that
if $k'=\Tvn{mate}(\wik)$ (evaluated before the call
has changed $k$ and~$A$)
is not $i$, then $k'$ must find a new mate.
Moreover, after the call $k'$ is to
the right of the barrier in the case of an
insertion and to the left of it in the case of a deletion.

Exclude the special cases identified above
by assuming that
$\{i,i'\}\cap\{\wik,k'\}=\emptyset$.
Then it can
be seen that all required changes to the matching
can be effectuated by matching $i'$ and $k'$,
which is what the procedure \Tvn{write} does.
In the case of an insertion, this makes $i$ strong,
which implies that $a_i$ can be set to $x$ simply
by executing $\Tvn{simple\_write}(i,x)$
at the very end.

In addition, with $\ell=\min\{i',k'\}$, it
must be ensured that the call does not change~$a_\ell$
except if $\ell=i$.
In the case of an insertion, $\ell=i'$, and
if $i'\not=i$, the mate of $i'$ switches from being $i$
to being $k'$, so that it suffices to
execute $\underline{A[k']}:=\underline{A[i]}$,
which happens in the procedure.
The same assignment is executed if $i'=i$,
in which case it is useless but harmless, given that it
takes place before the call $\Tvn{simple\_write}(i,x)$.
In the case of a deletion, $\ell=k'$.
Here the procedure plays it safe by remembering
the value of $a_{k'}$ before the call in a variable~$v$
and restoring $a_{k'}$ to that value
at the end, unless $k'=i$, via the call
$\Tvn{simple\_write}(k',v)$.
This is convenient because $a_{k'}$
is not stored in a unique way before the call.

At this point
$i$, $i'$ and $k'$ have been ``taken care of'',
but $\wik$ still needs attention.
In the case of a deletion, either $\wik=k'$ or
$\wik$ is weak, so nothing more needs to be done.
In an insertion, the procedure saves the original
value of $a_{\wik}$ in $u$ and restores it afterwards
through the statement $\Tvn{simple\_write}(k+1,u)$.
This is necessary and meaningful only if $\wik$ is strong.
If $\wik$ is weak, however, the effect of the
statement---except for the harmless
possible elimination
of a spurious matching edge---is canceled through
the subsequent assignment to
$\overline{A[k']}$ and $\underline{A[k']}$.

We still need to consider the special cases that
were ignored above, namely calls with
$\{i,i'\}\cap\{\wik,k'\}\not=\emptyset$.
These form part~(b)
of Figs.\ \ref{fig:2}--\ref{fig:5}.
In fact, the number of special cases is quite limited.
If $\wik$ is weak before an insertion
or strong before a deletion, it is unmatched.
Thus if $i=\wik$, we have $i=i'=\wik=k'$,
and $i'=k'$ implies $i=\wik$.
On the other hand, each of the statements
$i=k'$ and $i'=\wik$ implies the other one.
Thus there are two cases to consider:
(1) $i=i'=\wik=k'$
and (2) $i=k'\not=\wik=i'$.

In case~(1), all writing to $A$ happens
to $A[i]$.
For insertion, the execution of
$\Tvn{simple\_write}(i,x)$ at the very end
ensures the correctness of the call.
For deletion, the execution of
$\overline{A[k']}:=i'$ at the end ensures that
$i$ is unmatched, which is all that is required.
In case (2), after an insertion,
$i$ and $\wik$ are both to the right of
the barrier, $a_i$ and $a_{\wik}$ are both nonzero,
and the execution of
$\Tvn{simple\_write}(i,x)$ and
$\Tvn{simple\_write}(k+1,u)$
ensures that $A[i]$ and $A[\wik]$ have the correct
values after the call.
After a deletion, $i$ and $\wik$ are both
to the left of the barrier and $a_i=a_{\wik}=0$,
and the execution of
$\overline{A[i']}:=k'$ and $\overline{A[k']}:=i'$
in fact ensures that $i$ and $\wik$ are both unmatched,
which is all that is required.

Since all operations of the data structure have been
formulated as pieces of code without loops
and $b=O(w)$, 
it is clear that the operations execute in constant time.

\section{The Last Bits and Pieces}

\subsection{Reducing the Space Requirements}
\label{subsec:hiding}

The space requirements of the data
structure of the previous section can be reduced
from $2 b N+w$ bits to $2 b N+1$
bits, as promised in
Lemma~\ref{lem:main},
by a method of~\cite{HagK17,KatG17}.
First, $b$ is chosen to satisfy
not only $b\ge\log_2\! n$, but $b\ge 2\log_2\! n$,
which is clearly still compatible with $b=O(w)$.
As a result, for each $i\in V$ to
the left of the barrier, $A[i]$ has at least
$2\log_2\! n-\Tceil{\log_2\! N}\ge\Tceil{\log_2\! N}$ unused bits.
If $k\ge 1$, we store $k$ in the unused bits of~$A[1]$
(the unused bits of $A[2],\ldots,A[k]$ continue to be unused).
When $k=0$, even $A[1]$ is to the right of the barrier
and there are no unused bits in $A$, so we use a
single bit outside of $A$ to indicate whether $k$ is nonzero.
The resulting data structure occupies
exactly $2 b N+1$ bits.

\subsection{The Choice of $b$}

A practical choice dictionary
based on the ideas of this paper would probably
content itself with the main construction of
Section~\ref{sec:main} and refrain from
applying the construction of
Subsection~\ref{subsec:hiding} to squeeze
out the last few bits.
Then there is no reason to choose $b$ larger
than $w$, and $b=w$ seems the best choice.
This yields a self-contained atomic
choice dictionary that occupies $n+2 w$ bits
when used with universe size~$n$.

If $w$ is even and $w\ge 2\log_2\! n$,
another plausible choice is $b={w/2}$,
which allows an entry in the array $A$
to be manipulated with a single instruction
and simplifies the access to cells of~$A$.
It seems likely, however, that the gains in
certain scenarios from choosing $b={w/2}$
instead of $b=w$ are small and can be reduced
still further through an optimization of
the case $b=w$ that
omits superfluous operations on upper or lower
halves of cells in $A$.

If the space needed for an externally sized
choice dictionary is to be reduced all the
way to $n+1$ bits for universe size~$n$,
$b=2 w$ seems the best choice.

\subsection{A Self-Contained Atomic Choice Dictionary}

In order to convert the externally sized atomic
choice dictionary of Theorem~\ref{thm:main}
to a self-contained one,
we must augment the data structure with
an indication of the universe size~$n$.
This can clearly always be done with
$w$ additional bits.
If a space bound is desired that depends only on~$n$,
$n$ must be stored as a so-called
self-delimiting numeric value.
Assume first that the most significant bits in
a word are considered to be its ``first'' bits,
i.e., the ones to be occupied by a data structure
of fewer than~$w$ bits
(the ``big-endian'' convention).
Then one possibility is to use the
code $\gamma'$
of Elias~\cite{Eli75}:
With $\Tvn{bin}(n)$ denoting the usual binary
representation of $n\in\TbbbN$
(e.g., $\Tvn{bin}(10)=\texttt{1010}$),
store $n$ in the form of the string
$\texttt{0}^{|\mathit{bin}(n)|-1}\Tvn{bin}(n)$,
which can be decoded in constant time
with an algorithm of Lemma~\ref{lem:log}.
Since $|\Tvn{bin}(n)|=\Tceil{\log_2(n+1)}$,
this yields a space bound for the
self-contained choice dictionary
of $n+2\Tceil{\log_2(n+1)}$ bits.
If instead the least significant bits of a word are
considered to be its first bits
(the ``little-endian'' convention),
the scheme needs to be changed slightly:
The string $\texttt{0}^{|\mathit{bin}(n)|-1}\Tvn{bin}(n)$
is replaced by
$\widehat{\Tvn{bin}}(n)\texttt{0}^{|\mathit{bin}(n)|-1}$,
where $\widehat{\Tvn{bin}}(n)$ is the same as
$\Tvn{bin}(n)$, except that the leading \texttt{1}
is moved to the end.

Incidentally, if an application can guarantee
that $k$ never becomes zero, the method of
Subsection~\ref{subsec:hiding}
can be used to ``hide'' $n$
as well as $k$ in the array $A$
if we choose $b\ge 4\Tceil{\log_2(n+1)}$.
This yields a restricted self-contained
atomic choice dictionary that
occupies $n+1$ bits.
The restriction is satisfied, e.g., if
the universe $\{1,\ldots,n\}$ always contains
$4 b-1$ consecutive elements
that do not belong to the client set.

\subsection{Iteration}

We say that a choice dictionary with client set~$S$
supports iteration in constant time per element if
there are constant-time operations to reset the
iteration, to return (``enumerate'')
some ``next'' element of $S$,
and to test whether the iteration is complete,
i.e., whether all elements of $S$
have been enumerated.
A more precise description can be found in \cite{HagK16}.

Using Lemma~\ref{lem:log}, it is easy to enumerate
the positions of the nonzero bits
in a bit vector
of length $O(w)$ in constant time per position output---two
shifts can be used to clear the bits whose positions
were already enumerated.
This reduces the problem of supporting constant-time
iteration for the choice dictionary of
Theorem~\ref{thm:main} to that of supporting
constant-time iteration over
$\{i\in\{1,\ldots,N\}\mid a_i\not=0\}$
for the data structure of Lemma~\ref{lem:main}.
And the latter is easy:
For $i=k+1,\ldots,N$, enumerate $\Tvn{mate}(i)$.
The iteration needs to remember
$\Tceil{\log_2(n+1)}$ bits of state information
between calls.

\subsection{Additional Operations and Features}

Compared to the best previous choice
dictionary~\cite{HagK16}, the choice dictionary
of the present paper, while doing better on the
``core business'' of a choice dictionary, lacks many
important additional operations and features.
If an application needs any of these additional
capabilities, it must still resort to the
choice dictionary of~\cite{HagK16}.

First, the new choice dictionary supports
iteration over the client set $S$,
but it is not robust in the sense of \cite{HagK16}:
If $S$ is modified during an iteration through
insertions and deletions, the data structure will
not suffer a run-time error and will not enumerate
elements outside of $S$, but an element of $S$
may be enumerated more than once, and it may not
be enumerated at all even though it belongs to $S$
throughout the iteration.

Second, there is no obvious way to extend the new
dictionary to several \emph{colors}, i.e.,
to the maintenance of several pairwise
disjoint subsets of $U=\{1,\ldots,n\}$,
where $n$ is the universe size,
or even to support
the operation $\overline{\Tvn{choice}}$,
which returns an element in the complement of the
client set with respect to~$U$.

Operations that could be supported in constant time
are batched insertion and deletion
of all elements of a subset $I$ of $U$ with
$\max I-\min I=O(w)$ presented via a bit-vector
representation in a constant number of words,
as well as batched inspection,
in the sense of \Tvn{contains}, of $O(w)$
consecutive elements of $U$.
However, this is a characteristic shared with
earlier choice dictionaries.

\bibliography{all}

\begin{thebibliography}{10}

\bibitem{AngV79}
D.~Angluin and L.~G. Valiant.
\newblock Fast probabilistic algorithms for {H}amiltonian circuits and
  matchings.
\newblock {\em J. Comput. Syst. Sci.}, 18(2):155--193, 1979.

\bibitem{BanCR16}
Niranka Banerjee, Sankardeep Chakraborty, and Venkatesh Raman.
\newblock Improved space efficient algorithms for {BFS}, {DFS} and
  applications.
\newblock In {\em Proc. 22nd International Conference on Computing and
  Combinatorics ({COCOON} 2016)}, volume 9797 of {\em LNCS}, pages 119--130.
  Springer, 2016.

\bibitem{BriT93}
Preston Briggs and Linda Torczon.
\newblock An efficient representation for sparse sets.
\newblock {\em {ACM} Lett. Program. Lang. Syst.}, 2(1-4):59--69, 1993.

\bibitem{Eli75}
Peter Elias.
\newblock Universal codeword sets and representations of the integers.
\newblock {\em {IEEE} Trans. Inform. Theory}, 21(2):194--203, 1975.

\bibitem{ElmHK15}
Amr Elmasry, Torben Hagerup, and Frank Kammer.
\newblock Space-efficient basic graph algorithms.
\newblock In {\em Proc. 32nd International Symposium on Theoretical Aspects of
  Computer Science ({STACS} 2015)}, volume~30 of {\em LIPIcs}, pages 288--301.
  Schloss Dagstuhl -- Leibniz-Zentrum f\"ur Informatik, 2015.

\bibitem{FreW93}
Michael~L. Fredman and Dan~E. Willard.
\newblock Surpassing the information theoretic bound with fusion trees.
\newblock {\em J. Comput. Syst. Sci.}, 47(3):424--436, 1993.

\bibitem{Hag98}
Torben Hagerup.
\newblock Sorting and searching on the word {RAM}.
\newblock In {\em Proc. 15th Annual Symposium on Theoretical Aspects of
  Computer Science ({STACS} 1998)}, volume 1373 of {\em LNCS}, pages 366--398.
  Springer, 1998.

\bibitem{HagK16}
Torben Hagerup and Frank Kammer.
\newblock Succinct choice dictionaries.
\newblock {\em Computing Research Repository ({CoRR})}, arXiv:1604.06058
  [cs.DS], 2016.

\bibitem{HagK17}
Torben Hagerup and Frank Kammer.
\newblock On-the-fly array initialization in less space.
\newblock In {\em Proc.\ 28th International Symposium on Algorithms and
  Computation (ISAAC 2017), to appear}, 2017.

\bibitem{HagKL17}
Torben Hagerup, Frank Kammer, and Moritz Laudahn.
\newblock Space-efficient {E}uler partition and bipartite edge coloring.
\newblock In {\em Proc.\ 10th International Conference on Algorithms and
  Complexity (CIAC 2017)}, volume 10236 of {\em LNCS}, pages 322--333.
  Springer, 2017.

\bibitem{KamKL16}
Frank Kammer, Dieter Kratsch, and Moritz Laudahn.
\newblock Space-efficient biconnected components and recognition of outerplanar
  graphs.
\newblock In {\em Proc.\ 41st International Symposium on Mathematical
  Foundations of Computer Science ({MFCS} 2016)}, volume~58 of {\em LIPIcs},
  pages 56:1--56:14. Schloss Dagstuhl -- Leibniz-Zentrum f\"ur Informatik,
  2016.

\bibitem{KatG17}
Takashi Katoh and Keisuke Goto.
\newblock In-place initializable arrays.
\newblock {\em Computing Research Repository ({CoRR})}, arXiv:1709.08900
  [cs.DS], 2017.

\end{thebibliography}

\end{document}